\documentclass[prd,aps,tightenlines]{revtex4}

\usepackage{amsmath}
\usepackage{verbatim}
\usepackage{graphicx}
\usepackage[usenames]{color}
\usepackage{psfrag}
\usepackage{hyperref}

\def\beq{\begin{equation}}
\def\eeq{\end{equation}}
\def\bea{\setlength\arraycolsep{1.4pt}\begin{eqnarray}}
\def\eea{\end{eqnarray}}
\def\bse{\begin{subequations}}
\def\ese{\end{subequations}}
\def\bit{\begin{itemize}}
\def\eit{\end{itemize}}

\def\ubar{\_}

\begin{document}

\title{The Universe as an Inside-Out Star}
\author{Mitch Crowe, Adam Moss and Douglas Scott}
\address{School of Physics and
Astronomy, University of British Columbia, Vancouver, British Columbia,
 V6T 1Z1, Canada}
\date{\today}

\begin{abstract} 
Acoustic modes can be used to study the physics of the interior of a cavity,
and this is especially useful when the inside region is inaccessible.  Many
astrophysicists use such sound waves as an essential tool in their research.
Here we focus on two separate sub-fields in which oscillations on
the surface of a sphere are studied -- Helioseismology and CMBology -- the
surface being either the solar or cosmic photosphere.  Both research areas
use the language of spherical harmonics, as well as sharing many close
similarities in the underlying physics.  However, there are also many
fundamental differences, which we explain in this pedagogical article.
\end{abstract}
\maketitle


\section{Introduction}

`CMBology' (or CMB-Cosmology), the study of temperature fluctuations in the
Cosmic Microwave Background (CMB \cite{cmb}), and Helioseismology, the study
of acoustic oscillations on the surface of the Sun \cite{helio},
are two fields of much
experimental and theoretical interest today.  In the last decade, our knowledge
of both areas has increased dramatically through an active ground and
space-based observational programme.  This has led to substantial improvement
in the quantification of models for both the early Universe and the
interior of the Sun, and hence given us a deeper understanding of the
underlying physics.

These two areas are vastly different in terms of scale.  Firstly, consider the
difference in physical size.  Cosmology -- literally `the study of the whole
Universe' -- encompasses scales so vast as to make even our own Galaxy
seem minuscule.  The study of solar oscillations, on the other hand, is
constrained to an object barely 100 Earths in diameter -- virtually
non-existent on a cosmological map.  We can also consider the vast temporal
disparity between the two fields.  The photons we observe from the Sun describe
it as it was about 8 minutes ago, while CMB photons give an imprint of the
Universe as it was more than 13 billion years ago. 

Although the Sun and CMB are very different in terms of scale, the underlying
physics enabling us to understand them is essentially the same -- the physics
of sound waves resonating in a cavity.  Acoustic waves are a common everyday
physical phenomenon, and every undergraduate learns about the use of standing
sound waves to understand the interior of a cavity.  Probing the same kinds
of waves in the early Universe and solar interior is of great importance,
since they can be used to study the insides of objects which are otherwise
unreachable.

In this article we will compare the physics of these 2 astrophysical arenas:
CMB anisotropies and and helioseismology.  Both use similar language, talking
about acoustic modes, the photosphere and spherical harmonics, and hence it
should come as no surprise that there are very close physical analogues which
can be drawn.  However, as we will see, this is only possible
{\em if one thinks about the Universe as an inside-out version of a star}.


\section{The CMB and Helioseismology}

\subsection{Standard Cosmological Model}

All current cosmological data point towards the Hot Big Bang picture, in
which the early Universe was full of relativistic particles (baryons and
electrons) and radiation, along with components of dark matter and dark energy.
This extremely dense and hot Universe evolved by expanding and cooling -- as
does an expanding gas.  Most of the initial energy was in radiation, which
drove the expansion according to the equations of General Relativity.  Because
of the expansion, radiation from earlier times reaches us with stretched
wavelengths, and it is natural to use the observed redshift, $z$, as a label
for the epoch we are observing -- $z$ increases monotonically as we look
farther away and back to earlier times.

The immensely high temperatures meant that photons -- whose energy
distributions well approximated a black-body spectrum -- were originally
energetic enough
to keep the atoms ionized.  The primordial gas, then, must have been optically
thick, due to the high cross-section for Thomson scattering of the free
electrons.  Photons and baryons were said to be `coupled' during this era; the
continuous scattering off one another linked their temperature and fluid
properties.

As the Universe cooled, the rate of expansion fell due to the overall
gravitational attraction of matter.  A number of important epochs occurred as
particle interaction rates fell below the expansion rate.  One example is the
formation of the light elements at a temperature of around $1\,$MeV (redshift
of several billion), as the conversion between neutrons and protons
froze out.  This gave a helium mass fraction
of around 25\%, in fairly close agreement with the value in the solar
interior today.

As the Universe continued to cool, the ratio of the energy density in massive
particles relative to radiation increased, until the epoch of matter-radiation
equality was reached.  At a redshift several times smaller than this
($z\,{\simeq}\,1000$), very few photons were energetic enough to ionize
hydrogen, and so electrons were captured by protons, leading to an optically
thin universe -- a process referred to as cosmological recombination.  Photons,
having been freed from their electron captors travelled unhindered through a
mostly empty space until they are seen today as the CMB.  The probability for
a photon last scattering with an electron peaks at $z\,{\simeq}\,1100$, and so
we call this the `last scattering surface'.

Photons which last scattered at this epoch are seen today with a remarkably
pure black-body distribution at a cool temperature of about $2.7\,$K.
At a redshift of 1100 this means that recombination occurred at a temperature
of about $3000\,$K.  This cosmic photosphere is seen at an age of around
400{,}000 years, while the age of the Universe today is about 14 billion years.

The background radiation, as seen by any observer, is remarkably isotropic,
but contains the signatures of primordial structure in the form of temperature
anisotropies on the order of one part in $10^5$.  It is these anisotropies
which are of most interest to cosmologists, as their measurement promises
to constrain the many otherwise free parameters in theoretical models of the
evolution of the Universe.  Moreover, they are caused by the small fluctuations
in density which grow in contrast to become the rich structure
(galaxies, stars, people) that we see today.


\subsection{Standard Solar Model}

Not unlike the CMB, the solar surface is mostly uniform.  When we observe the
Sun, we see photons escaping from the solar photosphere, the energy
distribution of which is crudely that of a black-body with an effective
temperature $T\,{\simeq}\,5800\,$K.  Because of its brightness and angular
extent on our sky, solar temperature anisotropies were witnessed long before
scientific explanation could be provided for them.  Sunspots were first noted
by Galileo and could be seen with a tool as simple as a pin-hole camera.
However, these large solar `blemishes' are very localized on the solar surface,
as well as being transient, and as such do not contribute much to the overall
variance of angular anisotropies; in CMB language, sunspots are localized
highly non-Gaussian cold spots with amplitude $\Delta T/T \simeq 0.1$.

The configuration of the solar interior can be inferred by applying the
standard equations of stellar structure, which derive from the principles
of thermal and hydrostatic equilibrium.  These are complicated by details of
energy generation through nuclear processes and energy transport by radiation
and convection.  The convective zone is confined to the outer 30$\%$
of the solar radius, where no radiative transport occurs.
Energy generation is driven by the conversion of hydrogen to helium at
similar temperatures to what was achieved in the Universe when the primordial
helium was formed. 

It is the acoustic oscillations -- anisotropies discovered in 1960 by observing
Doppler shifts in absorption lines due to the physical movements of atoms in
the photosphere -- which are of greatest interest to helioseismologists.  These
oscillations, composed of various pressure modes or `p-modes', are waves
sustained by a radial pressure gradient; they are sound waves trapped in the
solar interior.  The principle underlying helioseismology is that the various
acoustic modes provide different information about the solar interior.
In particular, modes characterized by different numbers of radial nodes
penetrate to different depths within the Sun, providing a series of probes
allowing one to determine the radially dependent physics of the Sun.  For
example, by measuring the dispersion relation of a mode, one can estimate
the average sound speed it experiences.  Using several modes, and knowing their
penetration depths, a helioseismologist can determine the functional form of
the solar sound speed with respect to radius.  Such tests can be used to both
confirm and to constrain parameters within the Standard Solar Model, including
determining the interior composition and rotation rate.


\subsection{The Universe as an inside-out star}

Some similarities between the Sun and the Universe should already be
apparent.  Consider, for instance, that an observation of either the CMB or the
Sun collects photons originating from a spherical surface, and describing a
nearly uniform black-body spectrum.  In the Sun the photosphere is the surface
where gas density has increased sufficiently for photons to be strongly
scattered, and its radius is usually defined as $R_\odot$, which is around
$700{,}000\,$km.  In the CMB the last scattering surface has a distance from
the Big Bang which is given by the recombination time times the speed of
light.  {\it However}, the CMB sky surrounds us, and when we look out towards
the cosmic photosphere it is like looking {\it into\/} the surface of a star.

In contrast, the Sun is localized on our sky.  That
is to say, one can point a finger at the
centre of the Sun with the knowledge that it is entirely contained within
some finite radius of that point (at a given time).  This is not possible for
the CMB last scattering surface, where every observer (potentially in
quite different parts of the Universe) has their own
last scattering surface.  This is because in the uniformly expanding
Universe everything is moving away from everything else --
thus there is no true centre of the universal expansion or, rather,
every point can be considered to be the centre.  So every observer sees
the early Universe photons arriving from all directions
in a spherical shell around them.  We are at the centre of a space in which
the cosmic photosphere surrounds us, as if the surface of the star had been
wrapped all round us.  The centre of this `star' is then located in the very
early Universe, well beyond the distance of the last scattering surface,
and this centre (the position of the `Big Bang' if you like) is in every
direction as we look out, {\it into\/} the `cosmic star'.

\begin{figure}
\centering
\mbox{\resizebox{0.99\textwidth}{!}{\includegraphics{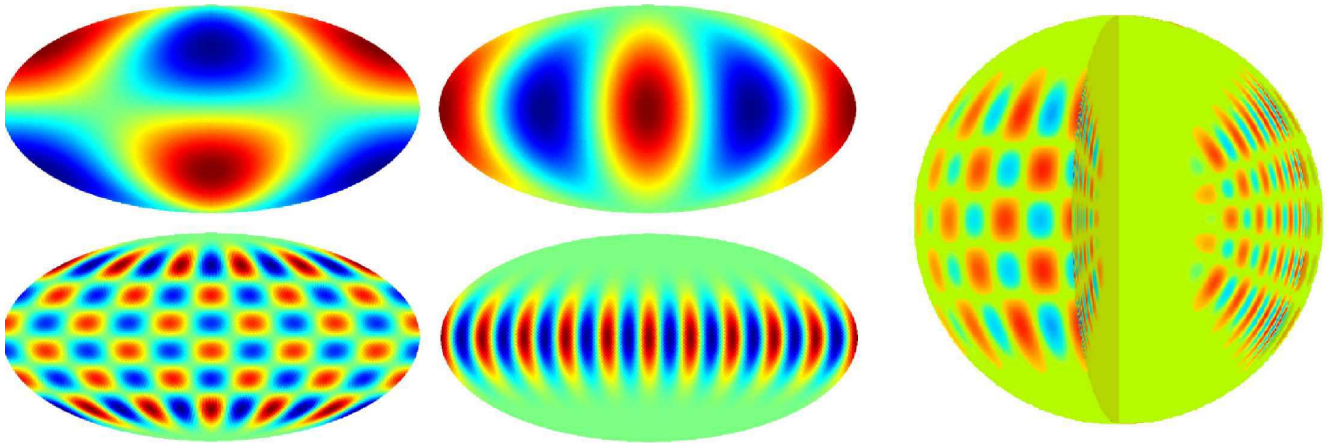}}}
\caption{\label{fig:harmmodes} Illustration of the $Y_{\ell m}$ harmonics used
to expand the temperature fluctuations of the CMB and velocity fluctuations on
the solar surface.  In the left panel of four harmonics, we show the real
component of: $Y_{2\,1}$ (top-left), $Y_{2\,2}$ (top-right), $Y_{10\,5}$
(bottom-left) and $Y_{10\,10}$ (bottom-right).  Cosmologists are interested in
the amplitude of these modes, while helioseismologists study their variation
with time.  Helioseismologists also decompose modes in the third (radial)
dimension, as shown on the right.}
\end{figure}

The Solar and Cosmic photospheres are therefore quite analogous to each other.
It is the fluctuations over each of these spheres which are of most interest,
since they probe the nature of the acoustic cavities.
Helioseismologists and cosmologists use the same mathematical tools to describe
these acoustic modes, namely the spherical harmonics.
Any angular function $f(\theta,\phi)$ can be
expanded in terms of spherical harmonics by
\beq
f(\theta,\phi) = \sum_{\ell=0}^{\infty}
 \sum_{-\ell}^{\ell} a_{\ell m} Y_{\ell m} (\theta,\phi)\,.
\eeq
Several of these modes are illustrated in Fig~\ref{fig:harmmodes}.  Roughly
speaking, the $\ell$ index describes the angular extent of features,
$\simeq180^\circ/\ell$, while $m$ characterizes the azimuthal dependence.
Cosmologists are mainly interested in the power spectrum of $a_{\ell m}$s,
while helioseismologists study their time dependence.  The fact that the
Universe is like an inside-out star means there is also another important
difference to keep clear.
The spherical harmonics describing anisotropies in the Sun and the CMB
must be projected from a different centre in each cases --
helioseismologists use the centre of the Sun as the origin, while cosmologists
use the observer's position as the centre.

\begin{figure}
\centering
\mbox{\resizebox{0.55\textwidth}{!}{\includegraphics{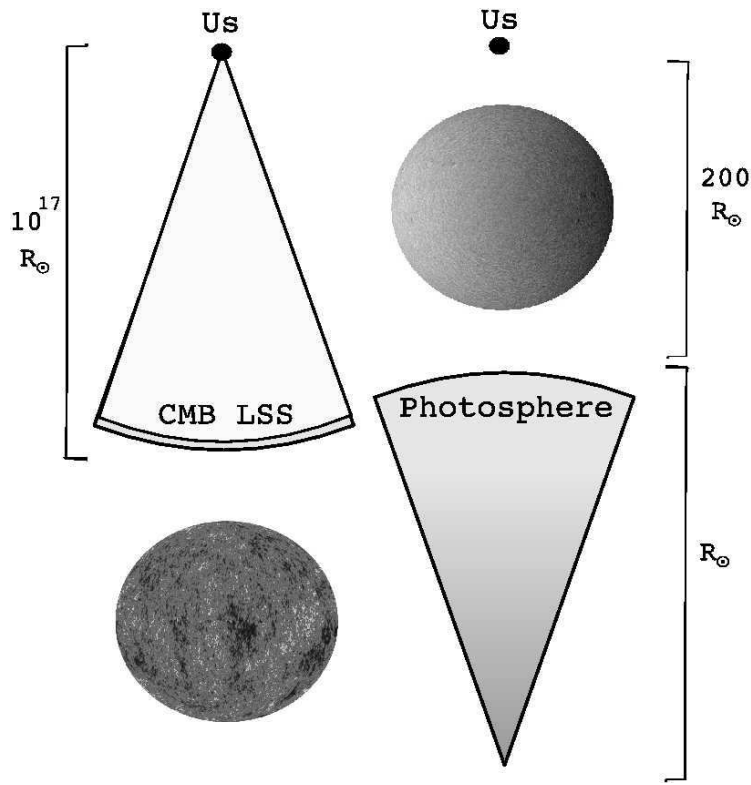}}}
\caption{\label{fig:inside_out} Geometry and length scales of the Sun
versus the comic inside-out star.}
\end{figure}

The scales and `inside-out' geometry are illustrated in
Fig.~\ref{fig:inside_out}.  The other major difference is the colossal distance
of the cosmic photosphere on which we see the CMB anisotropies.
At a redshift of around 1100, the last scattering surface is
about $14\,$Gpc away from us (a bit more than the light travel time in
14 billion years, because it has been expanding during that interval).
This is about $10^{17}$ times the scale of a Sun-like star.  So the Universe
is exactly like a star, except that it is completely turned inside-out,
and 100 quadrillion times bigger!

\subsection{Surface of last scattering and the photosphere}

The reason why the inside-out star is an attractive analogy for the CMB is that
we observe photons emerging from the cosmic photosphere.  This last-scattering
surface is a shell defined by the distance from us in which there is a
significant probability for the photons to have suffered their last scattering
event.  We see to where the optical depth $\tau$ is around unity, with $tau$
entirely due to Thomson scattering off free electrons.  The last scattering
surface is therefore defined by the epoch at which the Universe went from being
a plasma to being a neutral gas; the time since this epoch, coupled with the
finite speed of light, defines this spherical surface.  For the Sun (or any
other star) the photosphere is also defined by the region where
$\tau\,{\sim}\,1$, but the source of opacity is much more complicated,
including lines from heavy elements and $H^-$ ion scattering.  But there is
a more significant distinction, coming from the vastly different 
photon-to-atom ratio -- approximately 1 billion for the Universe, but
less than one {\it billionth} for the solar surface!  Hence the temperature at
which the Universe went from plasma to neutral is determined by when the
photons allowed the atoms to recombine, and this determines the
last-scattering surface.  But the photosphere of the Sun is mainly
determined by where the {\it density\/} has fallen off.  So it is much 
more like an actual edge to the solar material than in the case of the
Universe, where the density is slowly varying, but the ionization changes.

One might then ask why the cosmic recombination temperature turns out to be
within a factor of 2 of the temperature of the solar surface.  The answer is
that partly this was just a coincidence.  But the fact that the order of
magnitude is similar is {\it not\/} surprising, since this comes basically
from the temperature at which atoms get ionized.  So, despite the chemistry
being different, this is ${\sim}\,$eV in both cases.

Another important feature for the CMB is acoustic damping, which is related
to the thickness of the last scattering surface.  This turns out to be
approximately $\Delta z/z\,{\simeq}\,0.1$, which leads to a smearing of the
anisotropies at an angular scale which is about 10 times smaller than the
causal scale at the last scattering epoch, corresponding to
$\ell\,{\simeq}\,1000$, as can be seen in Fig.~\ref{fig:modes}.

Similarly, in the Sun the optical depth does not drop instantaneously, although
it much more abrupt than for the CMB.  The thickness of the solar photosphere
is a few hundred km or around $0.001\,R_\odot$.  This means that
the damping for solar modes is at an angular scale about 100 times higher in
$\ell$ than for the CMB.


\section{Physics of sound waves} \label{sec:sound}

\subsection{Acoustic modes}

A crucial property of both CMB and solar fluctuations is that their amplitude
is small.  As a result, the equations which describe them can be solved by
linear perturbation theory, such that a set of initial conditions can be
evolved forward in time exactly to predict the final observed oscillation
spectrum (particularly in the cosmic case), greatly simplifying the comparison
of theory with observations. 

Both the fluids in the solar interior and early Universe are characterized by
several variables, each with an average value and a small perturbation which
varies with both time and position -- the energy density $\rho$, pressure $P$
(related by an equation of state $P=P(\rho,T)$) and local velocity $v$.
A continuity equation enforces conservation of mass, Euler's equation
determines the motion of the fluid, and Poisson's equation describes the
response of the fluid to gravity.  The full cosmological equations are
relativistic generalizations of these, but the physics is the same. 

In order to understand acoustic waves it is sufficient to initially ignore
gravity and consider a density perturbation in each fluid.  The cosmological
fluid is adiabatic, a natural outcome of inflationary initial conditions, and
this is also an excellent approximation in most of the solar interior.  This
means that one can ignore heat exchange between fluid elements -- only
compression increases the temperature, and expansion cools it.  In both cases,
the resulting continuity and Euler equations then describe a simple oscillator 
\beq \label{eqn:sound}
\ddot{\delta} + \frac{\dot{a}}{a} \dot{\delta} + c_{\rm s}^{2} k^2 \delta =
 0\, ,
\eeq
where $\delta\equiv \delta \rho / \rho$, $\delta \rho \ll \rho$ and an overdot
denotes the time derivative.  The scale factor $a(t)$ describes the
cosmological expansion, and is related to redshift by $a=(1+z)^{-1}$ -- this
term is zero in the solar case.  We have written
this equation in Fourier space, where $k=2\pi/ \lambda $ is the wavenumber of
the mode.  Neglecting the expansion term, the solution to this equation are
plane acoustic waves oscillating at the sound speed, defined through
$c_{\rm s}^{2}=dP/d\rho$.  In a relativistic cosmological fluid
$c_{\rm s}^{2}=c^{2}/3$, where $c$ is the speed a light.  In the solar interior
the sound speed varies as a function of radius, and for an ideal gas we would
just have $c_{\rm s}^2=k_{\rm B} T/{\bar m}$, where $k_{\rm B}$ is Boltzmann's
constant and ${\bar m}$ is the average particle mass.  In the standard model
of the Sun the values range from about $6\,{\rm km\, s}^{-1}$ near the solar
surface to about $500\,{\rm km\, s}^{-1}$ near the solar core.

Equation~(\ref{eqn:sound}) has several important consequences \cite{acoustic}.
The dispersion relation of the waves, given by $\omega=c_{\rm s} k$, means that
spatial and temporal modes are related by a constant in the cosmological fluid
-- i.e.~a wave with twice the wavelength has twice the oscillation timescale,
which turns out to be crucial for understanding the
observed CMB acoustic spectrum.  In the solar interior, the oscillation
equations must be combined with boundary conditions at the surface.  Coupled
with the fact that the sound speed varies as a function of depth, this leads to
{\em refraction} of waves in the solar interior.  High frequency modes are
trapped near the surface, while low frequency modes penetrate closer to the
core.  This means that observation of oscillations as a function of frequency
can be used to probe the interior.

Since the oscillations in density in equation (2) also involve oscillations
in velocity, then there may be more than one physically observable
effect.  The velocities will be $90^\circ$ out of phase, since the velocity
maxima and minima occur at the zeros of the density oscillations.  For the
Sun it is usually the time-varying Doppler shifts of the velocity
oscillations that are observed directly (although there are also luminosity
variations observable from the lowest multipole modes).  For the CMB the
amplitudes of the standing waves are frozen at the last scattering epoch
-- the main effect is from the density variations, although there is a
sub-dominant contribution from the velocities.

Another important property of the acoustic modes is that they are
{\em irrotational}.  This means that the fluid velocity is in the
direction of the wavevector {\bf k}.  The cosmological fluid is dominated by
irrotational modes, which are the primary source of anisotropy in the CMB.
However, inflationary initial conditions also predict a small fraction of
rotational modes, which are seeded by gravitational waves.  The most
characteristic effect of these is through their effect on the pattern of
CMB polarization in which (through analogy with curl in electro-magnetism)
they are usually referred to as `B-modes'.  These have not yet been detected,
but are one of the main motivations for future CMB missions.  In the Sun there
is some evidence for `r-modes' in the photosphere, which are similar to the
Rossby waves seen in the Earth's atmosphere and oceans.  These are driven
by the Sun's rotation.  So, although there is a very loose analogy with the CMB
`B-modes', there is also a very fundamental difference -- there are extremely
strict limits on the rotation of the Universe, coming from the
non-observations of spiral-like patterns in the CMB anisotropies.


\subsection{Cosmological oscillations}

Inflationary cosmology predicts that quantum fluctuations created in the very
early Universe seeded gravitational fluctuations within the primordial plasma.
The introduction of gravity means that pressure and gravity are now two
competing forces.  The gravitational instability imposed on these perturbations,
along with the counter-acting radiation pressure from the energetic and
numerically dense photons (still highly coupled to the baryons) leads to
acoustic oscillations within this early plasma. 

After recombination, the photons' restoring radiation pressure no longer
drives these oscillations and the perturbations in structure are frozen-out
as the CMB is released.  On the largest scales we simply see a reflection
of the initial conditions in the gravitational potential.  This is the main
physical effect for angles larger than that subtended by the causal length
at last-scattering, which corresponds to about $2^\circ$ on the sky
(about the width of your thumb held at arm's length, or coincidentally about
4 times the diameter of the Sun).

On smaller scales, the temperature fluctuations are due to the density and
velocity variations in the oscillating plasma at the time of last scattering.
One observes these as a `fundamental' mode, together with a series of
harmonic overtones in angular scale on the sky.  This occurs because of the
approximately constant sound speed within the plasma up to decoupling.
There therefore exists a scale, which at last scattering had suffered
maximal compression -- so CMB photons have a maximum temperature variance
at this angular scale.  This fundamental mode for the CMB is the angle
subtended by the `sound horizon' at last scattering, i.e.~the maximum distance
sound could propagate since the initial fluctuations were laid down.  This
corresponds to an angular scale of around half a degree on the sky, or
$\ell\,{\simeq}\,220$.  We observe a set of harmonic overtones at
$\ell\,{\simeq}\,220\,(2n+1)$, corresponding to the modes which have undergone
further oscillations to reach maximal compression, and we observe peaks at
$\ell\,{\simeq}\,440\,(n+1)$, corresponding to maximal rarefaction modes
(with $n=0 \ldots \infty$ in each case).

\begin{figure}
\centering
\mbox{\resizebox{0.46\textwidth}{!}{\includegraphics{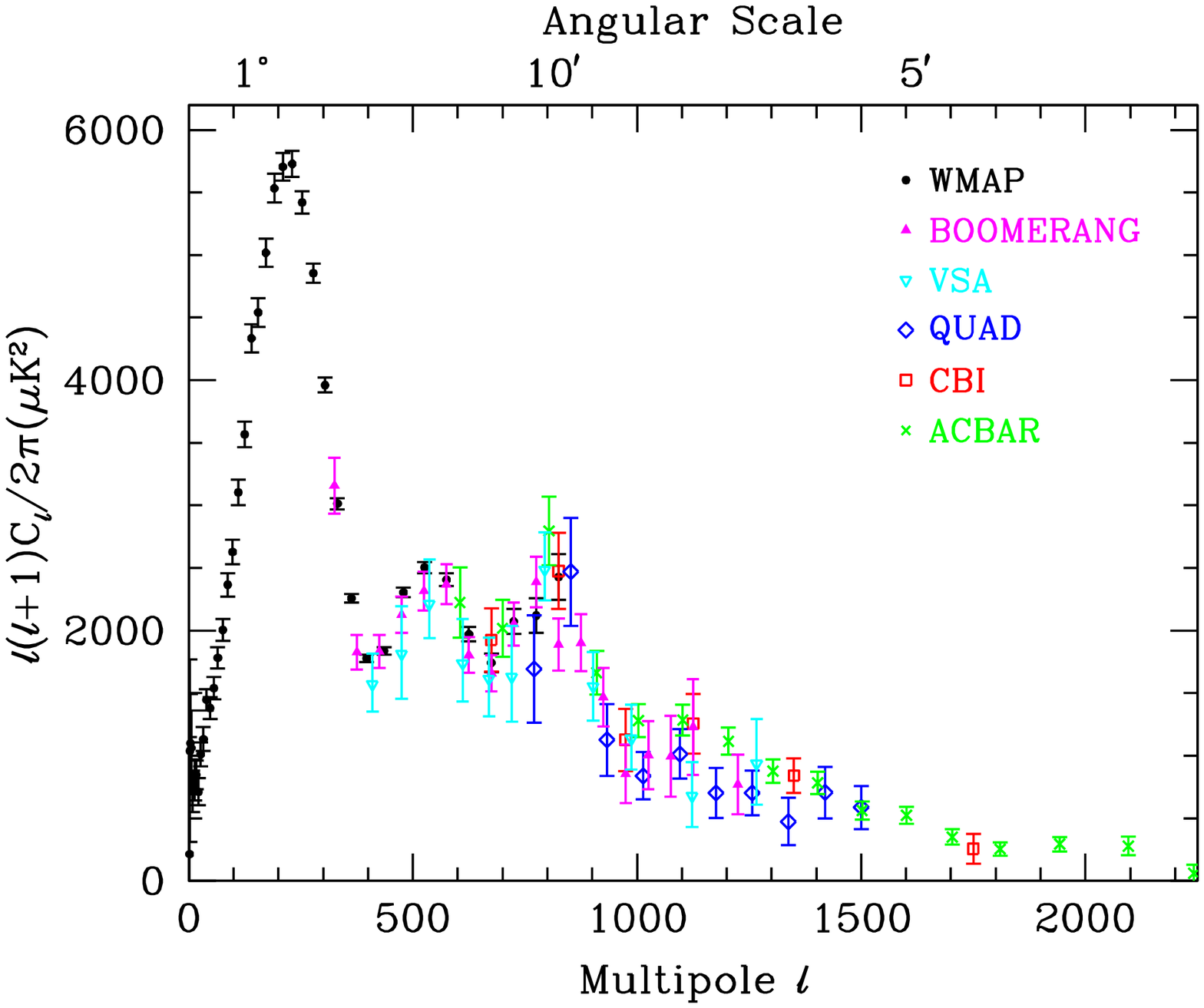}}}
\mbox{\resizebox{0.52\textwidth}{!}{\includegraphics{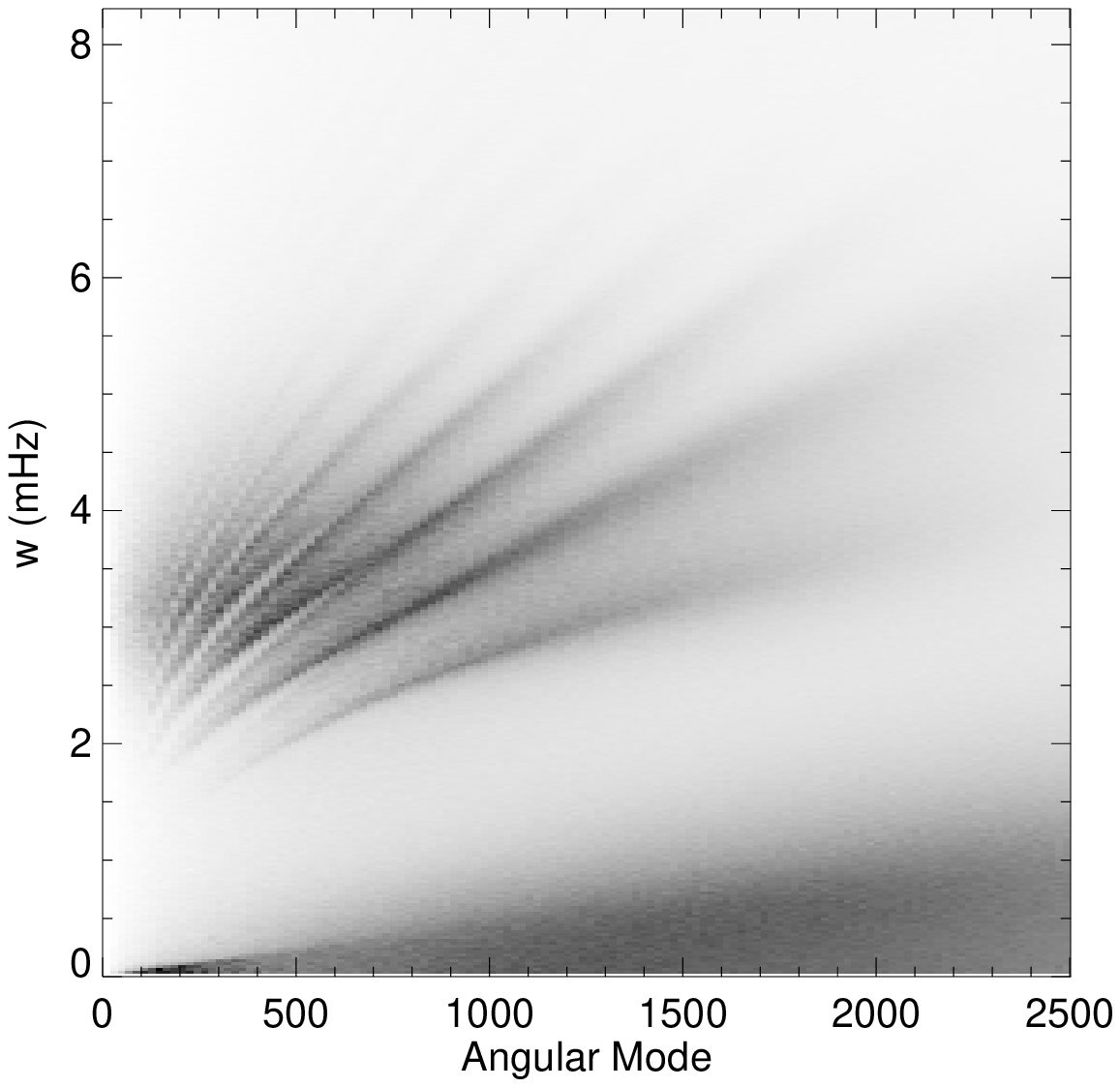}}}
\caption{\label{fig:modes} In the left panel, we show the CMB power spectrum
as measured by the latest data from the {\sl WMAP\/}~\cite{wmap},
BOOMERANG~\cite{boomerang}, VSA\cite{vsa}, QUAD\cite{quad}, CBI~\cite{cbi}
and ACBAR~\cite{acbar} experiments.  In the right panel, we show oscillation
modes of the solar surface, made using a representative example of
{\sl SOHO\/}~\cite{soho} MDI data.  The prominent ridges correspond
to different radial modes, while the low frequency structure is `noise'
below the sound speed.  The same diagram for the CMB would consist of a single
line, $\omega\propto\ell$, but with an extremely small coefficient, so
that it would lie almost along the $x$-axis.}
\end{figure}

These features are most easily seen by plotting the variance of CMB
temperature against angular scale on the sky, or more precisely by
plotting power versus multipole for spherical harmonics.  In the left panel of
Fig.~\ref{fig:modes}, we show the anisotropy power spectrum for current
observational data.  The conventional quantity which is plotted as power is a
scaled version of $C_{\ell}= \langle |a_{\ell m}|^2 \rangle$, where the angled
brackets mean an average over all possible realizations of each mode, and each
$m$ is equivalent (since there are no preferred directions in the Universe).
We see a clear detection of the first, second and third acoustic peaks.

This coherence of the CMB power spectrum only occurs because the initial
conditions are `synchronized'.  This happens naturally in inflationary
models for the initial perturbations when modes start oscillating at very
early times and over almost arbitrarily large scales (even those which are
apparently acausal).  This synchronization means that each cosmological
Fourier mode has the same temporal phase.  In the acoustic cavity analogy
this means that the modes have a node at $t=0$ (with the fundamental and
harmonics having nodes or anti-nodes at the recombination time).  This is
just like the radial modes in the Sun, which all have a node at the centre
and at the surface.  {\it However}, for the Sun the distance and epoch are
not tied together as they are for cosmological distance and look-back time.
The sound waves in the Sun may all have a node at the centre, but they are
excited at different (and random) times.

Inflationary initial conditions with amplitude about 1 part in $10^5$
appear to provide an excellent fit to today's cosmological perturbations --
they are approximately scale invariant in gravitational potential, are
adiabatic in nature and maximally random (i.e.~have Gaussian statistics, with
no phase correlations between modes).  This means that the fluctuations obey
purely Gaussian statistics,
so that the power-spectrum contains all useful cosmological information.
Of course this is manifestly not true for the Sun, or indeed any object where
one can meaningfully point at specific features.

\subsection{Solar oscillations}

In the CMB, the oscillation modes are mainly acoustic in nature.  In the Sun
the modes are acoustic (`p-modes'), gravity (`g-modes') and surface waves
(`f-modes').  Acoustic oscillations are driven by pressure variations, whereas
g-modes are driven by buoyancy of fluid parcels (gravity provides the restoring
force).  Surface waves arise from discontinuities in density along the surface
and propagate along these discontinuities.  Neither gravity nor surface waves
are possible in the CMB at linear order, due to the isotropy of temperature
at zeroth order.

In solar models spherical symmetry and boundary conditions at the surface are
enforced.  This leads to a set of discrete oscillation frequencies
$\omega_{n\ell m}$, where $\ell$ and $m$ are the degree and order of spherical
harmonics, and $n$ is the radial node number.  These frequencies are
independent of $m$ for a spherically symmetric non-rotating star.  However,
if the star is rotating (as is usually the case), this splits the frequencies,
in a similar fashion to the Zeeman or Stark effects for spectral lines.
The $+m$ modes travel at a
different speed around the star than the $-m$ modes.  Thus observations of this
splitting over many different multiplets enables reconstruction of the solar
interior rotation.

In the Sun the period of acoustic oscillation is dependent on the radial node
number $n$.  When a helioseismologist talks about the `fundamental',
what is meant
is the radial mode which has a node at the centre and an anti-node at the
surface.  The harmonics are then the radial modes with extra numbers of
nodes.  Each of these radial modes can have a whole set of angular harmonics
with different $\ell$ (and $m$).  The frequencies can be estimated from the
time taken for a sound wave to travel one horizontal wavelength, which
gives $\omega\,{\sim}\,c_{\rm s}\ell/R$ (for a wave propagating at depth $R$).
The fundamental node ($n=0$) has a period of ${\sim}\,1$ hour,
coming approximately from the free-fall time $(G\rho)^{-1/2}$.
Typical observed modes have $n=20$--30, with a period of ${\sim}\,5$
minutes.  However, there is not really a fundamental {\it angular\/}
mode for the Sun in the way that there is for the CMB
(although see Section~\ref{sec:sunpower}).

Since the solar oscillation timescale is much (much!) shorter than
the cosmological timescale, helioseismologists have a more powerful
signature of acoustics in the Sun than the acoustic peaks frozen onto
the CMB -- they can observe the {\em actual\/} oscillations and measure their
frequencies directly, instead of simply inferring the oscillations from a
harmonic imprint.  These observations can be captured in an $\ell$--$\nu$ plot,
where the angular frequencies are plotted against the temporal frequencies, as
shown in Fig.~\ref{fig:modes}.  The ridges in this plot correspond to node
number -- the ridge with lowest temporal frequency corresponds to the
fundamental mode.  Since modes with small $n$ probe deeper into the solar
interior, they can be used to constrain radially dependent properties.

Two key effects in the Sun prevent us from seeing coherent acoustic modes
like we observe in the CMB.  Firstly the
modes are probably generated by turbulent eddies in the convection zone, so
that there is stochastic forcing of the oscillations, with random temporal
phases.  And secondly the very long lifetime of a star compared with its
typical mode timescale ($\omega T_\odot\,{\sim}\,10^{15}$) ensures that even if
initial conditions were synchronized like for the Universe, stochastic
excitation would almost immediately lead to loss of coherence.

To summarize, in Table~\ref{tab:compare} we contrast some of the features of
the the Universe with those of the Sun.

\begin{table}[htdp]
\begin{center}
\begin{tabular}{ll}
 The Sun  & The Universe \\ \hline
 We're outside, looking in & We're at the centre! \\
 Solar photosphere & Last scattering surface \\
 ${\sim}\,0.1$\% thick photosphere & ${\sim}\,$10\% thick photosphere \\
 Photons made in nuclear core & Photons made in early nuclear processes \\
 Complex scattering & Thomson scattering \\
 Solar spectrum of absorption lines & Weak emission lines of H and He \\
 $T\,{\sim}\,6000\,$K & $T\,{\sim}\,3000\,$K (redshifted) \\
 G dwarf star & M (super-duper) giant star \\
 $R_\odot$ & $10^{17}R_\odot$ \\
 Helioseismic modes & CMB anisotropies \\
 Information from mode frequencies & Information from angular power spectrum \\
 Time variations ${\sim}\,5$ minutes
  & Time variations ${\sim}\,10^{10}$ years \\
 Stochastic, continual excitation & Synchronized initial conditions \\ 
 Rotation axis defines $m$ &  No preferred directions \\
\end{tabular}
\end{center}
\label{tab:compare}
\caption{Summary of similarities and differences between CMBology and
helioseismology.}
\end{table}%


\subsection{Observation and interpretation of oscillations}

At present, the instrument which has produced the most sensitive full-sky maps
of the CMB is the Wilkinson Microwave Anisotropy Probe ({\sl WMAP\/})
\cite{wmap}.  For
the Sun, the analogous instrument is the Michelson Doppler Imager (MDI) on
board the Solar and Heliospheric Observatory ({\sl SOHO\/}) \cite{soho}.

{\sl WMAP\/} measures temperature differences on the sky with an instrument
located at the L2 Lagrange point of the Earth-Sun system, which lies about
$1.5$ million km on the opposite side of the Earth from the Sun.
{\sl WMAP\/} has an angular resolution of $0.22^\circ$, which means it can
measure anisotropies up to $\ell\,{\sim}\,800$. 

MDI measures the Doppler shifts of the gas in the solar photosphere on a
spacecraft located at the L1 Lagrange point, which lies at the same distance
from the Earth as L2, but on the Sun side.  MDI has an angular
resolution of about $0.0006^\circ$ when observing the full solar disk, such
that the pixel resolution on the solar surface is around $R_\odot/500$.
However, since the Sun is {\it not\/} an inside-out star, the origin for
the spherical harmonic coordinate system is the solar centre, and this means
that MDI can measure anisotropies out to $\ell\,{\sim}\,1000$.  This is
coincidentally similar to the {\sl WMAP\/} resolution.

The solar oscillation spectrum in Fig.~\ref{fig:modes} was produced using 11
hours of nearly continuous high-resolution MDI solar dopplergrams,
consisting of one-minute integrations.  Helioseismologists studying such
data have access to several years of uninterrupted dopplergrams, but the basic
information is evident using this limited data-set.

The information extracted from the modes has a different form for each of the
2 fields we are comparing.  In cosmology one makes a CMB map, extracts the
power spectrum of anisotropies and typically uses a least-squares
routine to fit a cosmological model to the data.  It is remarkable that only 6
parameters (within a simple isotropic, homogeneous framework) are needed to
provide an excellent fit to {\sl WMAP\/} data.
In helioseismology, fitting to the  acoustic frequencies are often computed
using direct inversion techniques on the data.  These are then used to tune
the solar model, where the `parameters' include some unknown radial functions.


\section{A power spectrum of the Sun} \label{sec:sunpower}

CMBologists learn about the Universe from the CMB
power spectrum, because it contains almost all the useful information,
together with the fact that the linear perturbation theory is remarkably
simple.  In contrast, the full theory for the amplitudes of helioseismic modes 
would have to be {\it non}-linear, and hence is far from simple.  Because
of this, the amplitude information of solar modes is often set aside in order
to focus on the frequencies.

Nevertheless, we can ask what the mode power spectrum would look like
for the Sun, in analogy with the CMB $C_\ell$s.
In the solar $\ell$--$\nu$ plot (Fig.~\ref{fig:modes}) the ridges are
p-modes with different radial mode number, $n$, while the signal at
lower temporal frequencies is dominated by convective and other noise
effects.  This is mainly from convective granulation and supergranulation
motions, and although these are interesting in their own right, they will
obscure the acoustic oscillations.  We therefore remove this low (temporal)
frequency signal from the MDI data before attempting to make a power spectrum.
We do this by cutting out everything in $\ell$--$\nu$ which would lie
below the surface sound speed (following~\cite{Georgobiani}), which is
the lowest speed at which acoustic modes can propagate in the interior.

\begin{figure}
\centering
\mbox{\resizebox{0.7\textwidth}{!}{\includegraphics{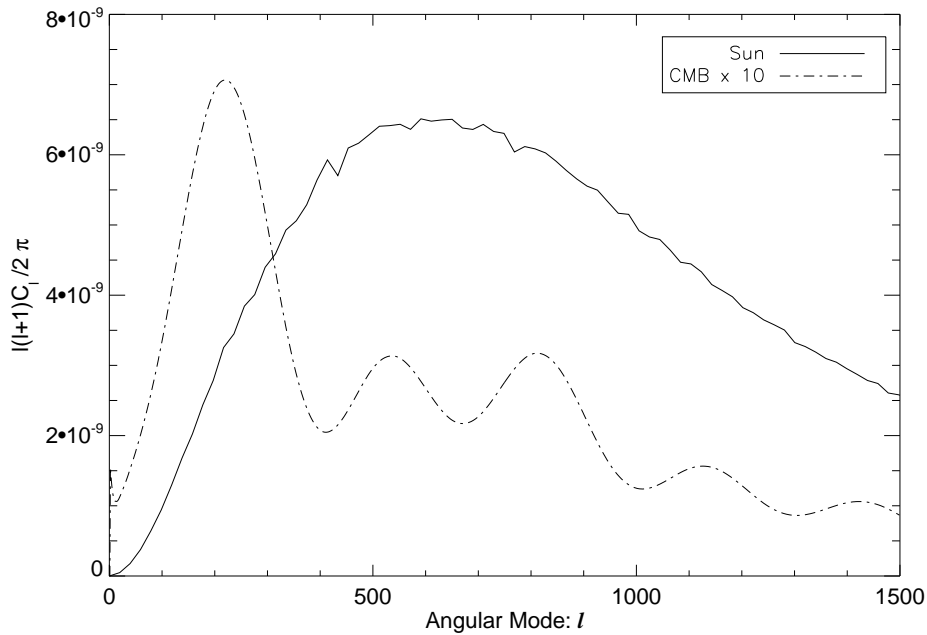}}}
\caption{\label{fig:powcomp} Comparison of CMB and solar power spectra.}
\end{figure}

We can make an order of magnitude conversion to temperature using a blackbody
model for the luminosity:
$L=4\pi R^2 \sigma T_{\rm eff}^4$, where $\sigma$ is the Stefan-Boltzmann
constant.  If for simplicity we take $L$ as constant over any oscillation,
then $\Delta T/T\,{\simeq}\,V/2\omega R_\odot$, where $V$
is an average of the velocity integrated over time, and $\omega$
is the measured oscillation frequency.
Performing this scaling to `temperature units', integrating MDI data
over all frequencies, and correcting for some instrumental efficiency
effects, gives the
curve shown in Fig.~\ref{fig:powcomp}.  Remarkably, the CMB power spectrum
needs only to be scaled up by about an order of magnitude in order to be
comparable.  This mean that in terms of $\Delta T/T$ amplitude the two power
spectra are within about a factor of three, although of course the shapes
of the 2 curves are quite different.

\subsection{Conclusions}

We have shown interesting analogies between the fields of CMBology and
helioseismology.  One could contrast more features -- for example looking at
the polarization information or comparing details of how the Sun and the
Universe have changed over time.  However, we have probably carried this
analogy far enough to be useful in understanding more of the physics of
both the CMB and the Sun.

As a final remark, we note that as well as learning about the Sun through its
acoustic structure,
astrophysicists are also beginning to learn about the interiors of other
stars -- the science of Asteroseismology.  For example, since 2003
the microsatellite {\sl MOST\/} \cite{most} has been studying low angular
degree modes in many nearby stars.  If there
is a further analogy to be drawn here, it may be that each of these stars
is like a separate inside-out universe, and Asteroseismology is like studying
the multiverse!


\section*{Acknowledgments} This research was supported by the Natural Sciences
and Engineering Research Council of Canada.  We thank Chris Cameron, Mark
Halpern, Jaymie Matthews and Jim Zibin for very useful conversations.

\end{document}